\documentclass[cits]{PoS}

\usepackage[utf8]{inputenc}

\title{Spinning-up: the case of the symbiotic X-ray binary 3A~1954+319 }

\ShortTitle{Spinning-up: the symbotic X-ray binary 3A 1954+319}

\author{\speaker{F.~Fürst}$^{a}$, D.~M.~Marcu${^b}$, K.~Pottschmidt${^b}$, V.~Grinberg$^a$, J.~Wilms$^{a}$,
 and M.~Cadolle Bel$^c$ \\
\llap{$^a$}Dr. Karl Remeis-Sternwarte \& ECAP, University Erlangen-Nuremberg, Bamberg, Germany\\
\llap{$^b$}CRESST/NASA-GSFC, Greenbelt, MD, USA \& UMBC, Baltimore, MD, USA\\
\llap{$^c$}ESAC, Madrid, Spain
}

\abstract{
We present a timing and spectral analysis of the variable X-ray source \ta. Our analysis is
mainly based on an outburst serendipitously observed during \inte Key Program observations of
the Cygnus region in 2008 fall and on the \swift/BAT longterm light curve. Previous observations, though
sparse, have identified the source to be one of only nine known symbiotic X-ray binaries, i.e., systems
composed of an accreting neutron star orbiting in a highly inhomogeneous medium around an M-giant
companion. The spectrum of \ta above $>20$\,keV  can be best described by a broken power law model. The extremely long
pulse period of $\sim$5.3 hours is clearly visible in the \inte/ISGRI light curve and confirmed through
an epoch folding period search. Furthermore, the light curve allows us to determine a very strong
spin up of $-2 \times 10^{-4}$\,\nicefrac{h}{h} during the outburst. This spin up is confirmed by the pulse period evolution
calculated from \swift/BAT data. The \swift/BAT data also show a long spin-down trend prior to the 2008
outburst, which is confirmed in archival \inte/ISGRI data. We discuss possible accretion models and
geometries allowing for the transfer of such large amounts of angular momentum and investigate the harder spectrum of this outburst compared to previously published results.}

\FullConference{8th INTEGRAL Workshop “The Restless Gamma-ray Universe”- Integral2010,\\
		September 27-30, 2010\\
		Dublin Ireland}

\usepackage{amssymb}
\usepackage{amsmath}
\usepackage{txfonts}
\usepackage{graphicx}
\usepackage{xspace}
\usepackage{multicol}
\usepackage{subfigure}
\usepackage{wrapfig}
\usepackage{microtype}
\usepackage{paralist}

\usepackage{nicefrac}

\newcommand{\chandra}{\textsl{Chandra}\xspace}

\newcommand{\ginga}{\textsl{Ginga}\xspace}
\newcommand{\inte}{\textsl{INTEGRAL}\xspace}
\newcommand{\sax}{\textsl{BeppoSAX}\xspace}

\newcommand{\swift}{\textsl{Swift}\xspace}

\newcommand{\msun}{\ensuremath{\text{M}_{\odot}}\xspace}

\newcommand{\redchi}{\ensuremath{\chi^{2}_\text{red}}\xspace}

\newcommand{\ta}{3A~1954+319\xspace}
\newcommand{\citep}[1]{\cite{#1}}
\newcommand{\citet}[1]{\cite{#1}}

\begin{document}

\section{Introduction}

\ta belongs to a small class of Low Mass X-ray Binaries (LMXB), in which a neutron star orbits around an M-type companion \cite{masetti06a}. The orbits are typically wide, making accretion via common Roche-Lobe overflow impossible. Instead, material from the wind of the companion is accreted onto the neutron star and thereby converted into X-rays. This accretion mechanism is very common among High Mass X-ray Binaries and their O- and B-type companions with strong stellar winds, but not well investigated for late type binary systems. In the case of \ta the accretion stream onto the neutron star seems to be highly variable as irregular X-ray flares are observed as well as times where the source is hardly detected at all. Binaries hosting a neutron star and an M-type companion are commonly referred to as Symbiotic X-ray Binaries (SyXB), in analogy of Symbiotic Stars consisting of a White Dwarf and a late-type star. These systems were named after their optical spectra in which both components, the hot White Dwarf and the cold companion are seen. Such features are not seen in SyXB, as the neutron star is invisible in the optical. A close look in X-rays is thus indispensable for a description of those systems.

Clearly confirmed as SyXBs are only 7 sources today \citep{nespoli09a}, two more are under dispute. Only very few of the SyXB, like GX~1+4 or 4U~1700+24 have already been studied in detail with different X-ray instruments, including high-resolution CCD detectors, and much is still to  be learned about these systems and the properties of the wind and the accretion mechanism. Being easily observable in X-rays SyXB provide a unique opportunity for studying the wind properties of late type stars, via X-ray variations of the emission lines and absoprtion column.

\ta was detected in the late 1970s with \textsl{Uhuru} and \textsl{Ariel V} \citep{forman78a,warwick81a}. \citet{tweedy89a} analyzed \ginga data and proposed that the system could be a HMXB. No more detailed studies were performed until \citet{masetti06a} identified the companion as an M-star using an accurate \chandra-position, placing the system in the group of SyXB. Soon thereafter, \citet{corbet06a} discovered a $\sim$5\,h period in \swift/BAT data. This lead to closer investigations by \citet{mattana06a}, who presented the first broad-band spectrum of \ta, using \inte and \sax. More spectral analysis was presented by \citet{masetti07b}, who also showed that the spectrum is consistent with the typical spectrum expected from an accreting neutron star and can be modelled using an either a cutoff powerlaw model or a thermal Comptonization model.

A strong decrease of the pulse period from 5.17\,h to 5.09\,h during an outburst in 2004/2005 and possible physical explanations thereof was reported by \citet{corbet08a} in \swift/BAT data. The period itself is most likely the spin period of the neutron star, as it is much too short for the expected wide orbit of the system and no other significant periods have been found between $5\times10^{-3}$ -- $3.5\times10^{+8}$\,s \citep{corbet08a, masetti07b}. It is also unlikely that the observed decrease in the period is due to the orbit of the binary system, as the mass function would then lead to impossible values of the mass of the companion, assuming a canonical mass of 1.4\,\msun for the neutron star.

In 2008 November \inte serendipitously observed a strong outburst of \ta, obtaining high quality lightcurves and spectra, which are presented here.

\section{Lightcurves and Timing}

\begin{figure}
\begin{minipage}{0.48\textwidth}
 \includegraphics[width=1.0\textwidth]{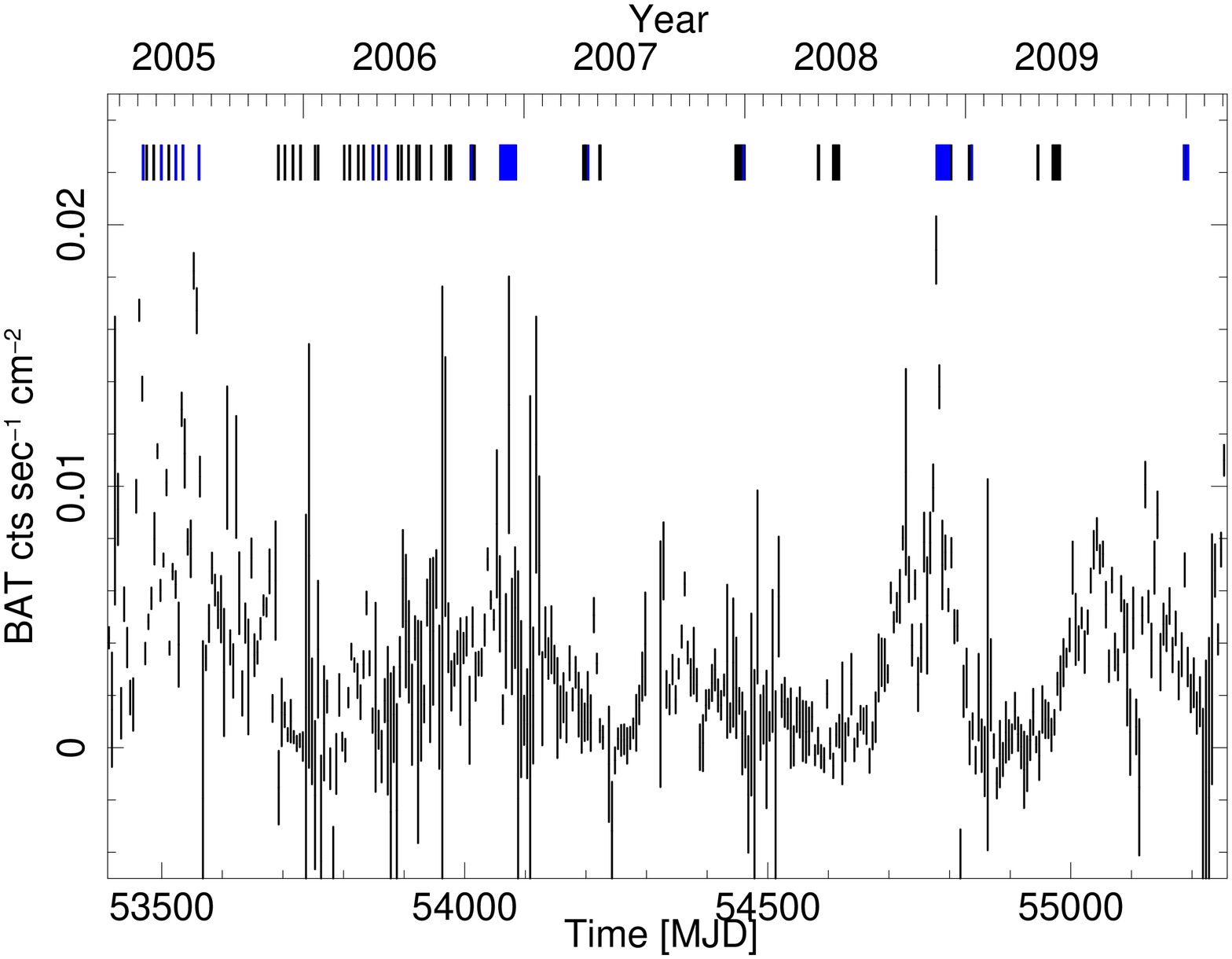}
 \end{minipage}
\hfill
\begin{minipage}{0.48\textwidth}
 \includegraphics[width=1.0\columnwidth]{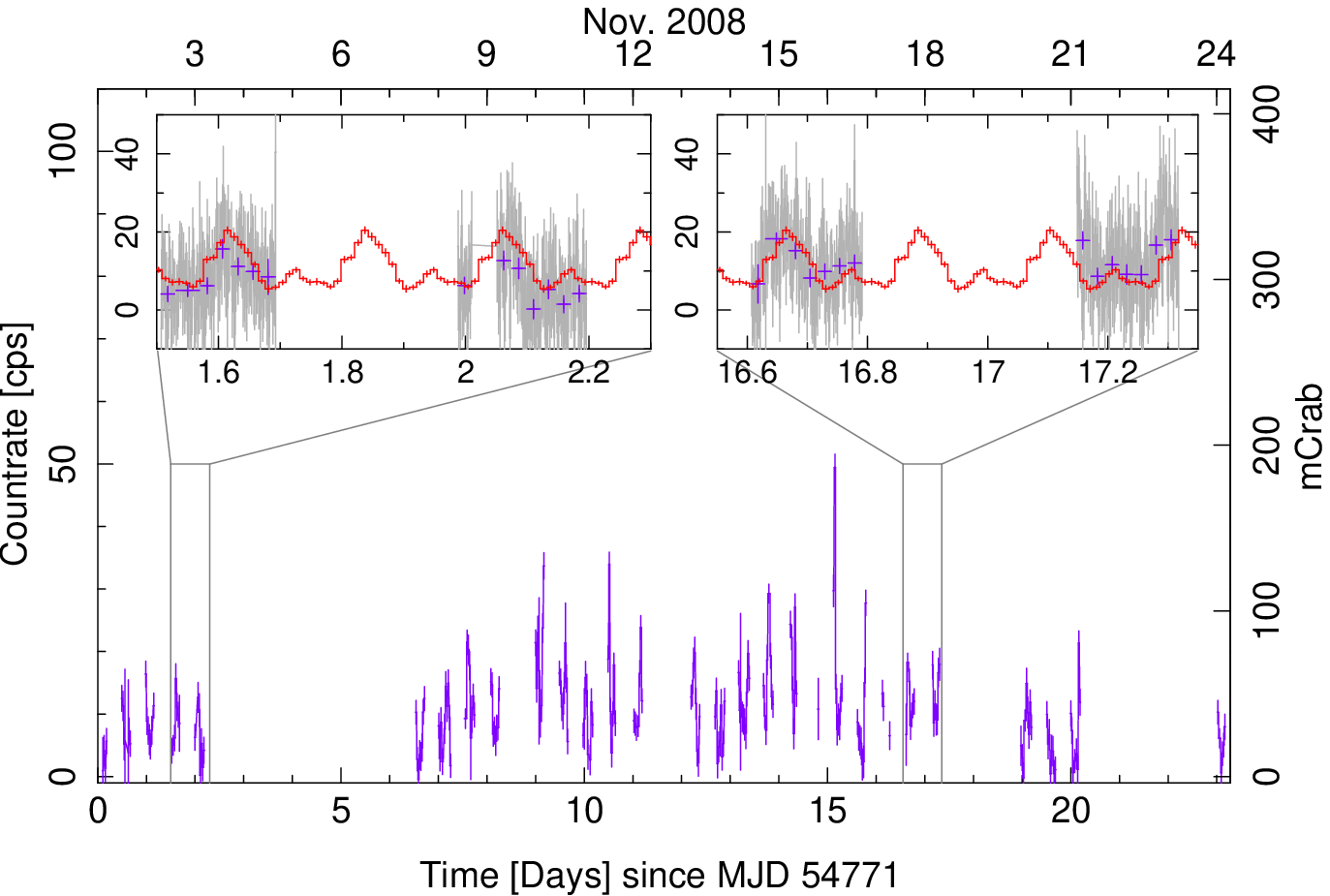}
\end{minipage}
 \caption{\textit{Left:} Lightcurve as measured with \swift/BAT. The tickmarks above indicate times of \inte observations, with the blue tickmarks showing times when the source was significantly detected in \inte/ISGRI. \textit{Right:} \inte/ISGRI 20--100\,keV lightcurve in November 2008. The data were binned to Science Window resolution. The insets show zooms  and a more detailed lightcurve with 100\,s resolution. Overplotted in red is the average pulse profile.}
\label{fig:batintlc}
\end{figure}

The long-term daily lightcurve of \ta as measured with \swift/BAT  is shown in Fig.~\ref{fig:batintlc}, left. It is clearly seen that \ta shows irregular flaring with some flares being as bright as almost 100\,mCrab and other episodes where it falls below the detection limit of \swift/BAT. Tickmarks above the lightcurve mark times when \ta was in the field of view of \inte/ISGRI. A relatively dense sampling of \inte observations was performed during a flare in 2008 November when Key Programme observations of Cygnus X-1 (PI Wilms) were scheduled. As seen in the left panel of Fig.~\ref{fig:batintlc} this flare was one of the brightest since the monitoring of \swift/BAT, with an average flux over the outburst of $\sim 8.6\times10^{-10}\,\mathrm{erg}\,\mathrm{cm}^{-2}\,\mathrm{s}^{-1}$ between 20--100\,keV, i.e., $\sim$60\,mCrab and a peak flux of more than 180\,mCrab. We extracted \inte data of that outburst using the standard Off-line Science Analysis (OSA) 8.0 to obtain spectra and using \texttt{ii\_light}, as distributed with OSA 7.0, to obtain lightcurves with a time resolution of 100\,s. The \inte/ISGRI lightcurve of that period is shown in Fig.~\ref{fig:batintlc}, right. Even though \ta was in an overall bright state during that time the \inte data show that the X-ray flux is still highly variable and varies by a factor of $\sim$20. Using the epoch-folding technique \citep{leahy83a}, a period of $\sim$5.3\,h is clearly detected in the data. Compared to the results by \citet{corbet08a}, the pulse period is significantly longer, i.e., the neutron star has spun down between 2005 and 2008.

\begin{figure}
\begin{minipage}{0.48\textwidth}
 \includegraphics[width=1.0\textwidth]{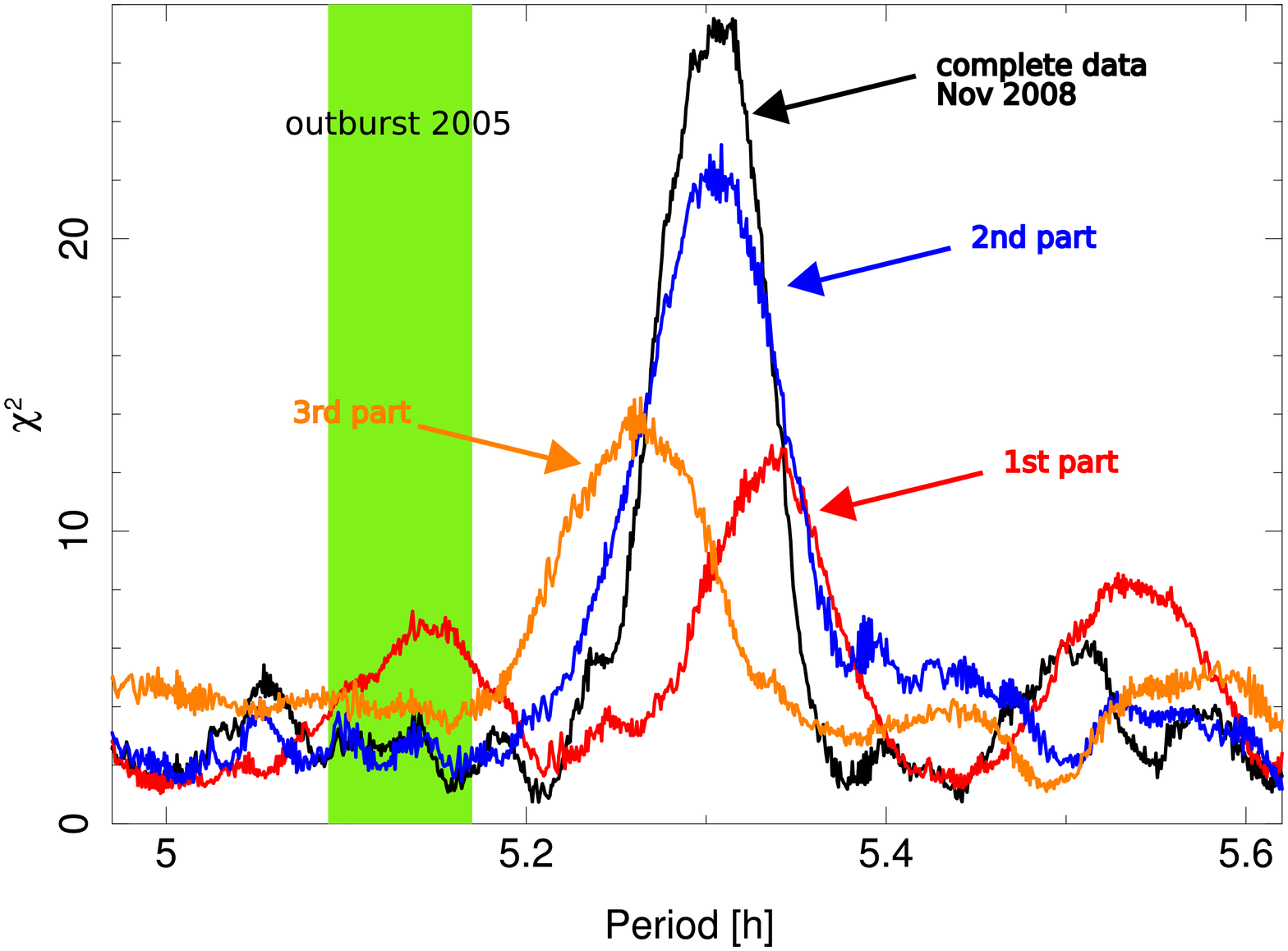}
 \end{minipage}
\hfill
\begin{minipage}{0.48\textwidth}
 \includegraphics[width=1.0\columnwidth]{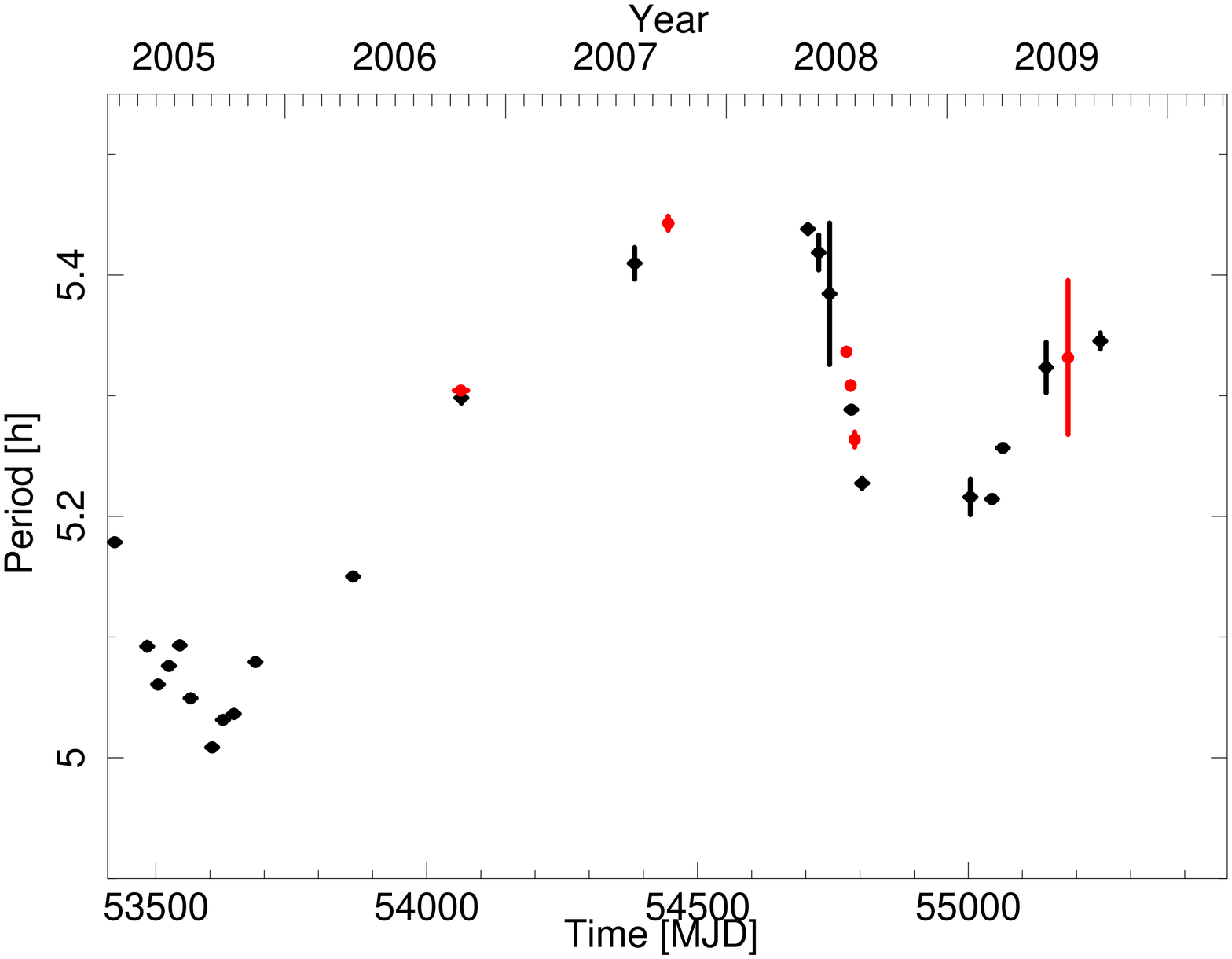}
\end{minipage}
 \caption{\textit{Left:} Results of the epoch folding analysis, with the lightcurve separated into three equally long parts. Additionally the overall result is shown in black and the range of the 2005 pulse period is marked in green. \textit{Right:} Long term pulse period evolution, measured with \swift/BAT (black) and \inte/ISGRI (red). }
\label{fig:epfo}
\end{figure}

As \citet{corbet08a} found a strong spin up trend in the 2005 flare, we split the \inte lightcurve into three equally long parts and searched in each one for the pulse period separately. We found a clear spin up trend in these data, with the pulse period decreasing from 5.34\,h in the beginning of the outburst to 5.26\,h in the end. The results of the epoch-folding are shown in Fig.~\ref{fig:epfo}. From this analysis we determined the pulse ephemeris to be (during that outburst): $T_0^\text{MJD} =54782.6897$\,d (mean of observation), $P= 5.3085\,\text{h}$, and $\dot{P} = -2 \times 10^{-4}$\,\nicefrac{h}{h}. To check for consistency, we folded the lightcurve with this ephemeris to obtain a pulse profile and compared it to the measured lightcurve. As seen in the insets of the right panel of Fig.~\ref{fig:batintlc} the pulse profile describes the lightcurve very well, in the beginning of the outburst as well as close to the end. This is also the first time that the pulsed flux of \ta is directly visible in a lightcurve, showing that \ta does not show large pulse-to-pulse variations.

 In order to put our \inte results into perspective,
we updated the \swift/BAT period analysis of \ta,
again using the epoch-folding technique. Using all publicly available \swift/BAT lightcurve data, local period determinations were performed for 20\,d long segments. Figure~\ref{fig:epfo} shows all successful \swift/BAT period measurements, that is, all data in which a period could be significantly measured. As the \swift/BAT data a impaired by a low signal-to-noise ratio, we also analyzed more \inte/ISGRI archive data. Additionally to the three data points from the 2008 November outburst we could obtain 3 more measurements of the pulse period during times with extensive coverage of \ta with \inte. All data points, \swift/BAT and \inte/ISGRI, are plotted together in the right panel of Fig.~\ref{fig:epfo}. As can be clearly seen the measurements for both satellites agree very well. The pulse period is slowly decreasing during quiescence but increases strongly during prominent outbursts, as expected from standard accretion theory \citep{ghosh79a}. One such outburst was measured by \citet{corbet08a} in 2005, the other one was caught serendipitously by \inte in 2008 November. Even though in between these outbursts \ta shows smaller flares, no measurable spin up is seen there. Correlating the luminosity with the change in pulse period led to no conclusive result, however, this may be due to the sparse sampling of the pulse period between outbursts.

\section{Spectral analysis}

\begin{figure}
 \centering
 \includegraphics[width=0.5\columnwidth]{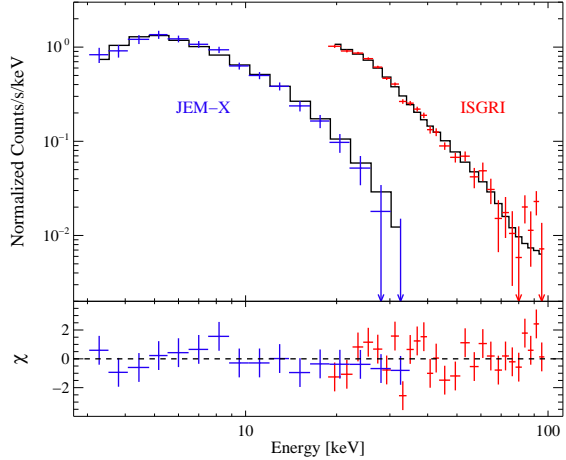}
 \caption{\textit{Top:} Spectrum taken with JEM-X and ISGRI, together with the best fit model. \textit{Bottom:} Residuals of the best fit.}
 \label{fig:spectrum}
\end{figure}

The \inte observations present the unique possibility to gain knowledge about the accretion physics during a strong flare of \ta, which was accompanied by a strong spin up of the neutron star. Due to the fact that the source is close to \inte/ISGRI's detection limit during pulse minimum, we created a peak spectrum of only those 50\% of data, which were measured at phases close to the major peak of the pulse profile, thereby improving the signal-to-noise ratio. Comparing pulse profiles at different energies above 20\,keV did not reveal any strong energy dependence with pulse phase, indicating that the spectrum is not sensitive to the selected pulse phase  and that the obtained spectrum is very similar to an overall phase averaged spectrum. This filtering resulted in the selection of 75 Science Windows (ScWs) with a pointing offset of $<10^\circ$ to \ta, leading to a pulse peak spectrum with 445\,ks exposure in ISGRI and 22\,ks in JEM-X. Modelling these JEM-X and ISGRI data between 3--100\,keV with the published models lead to unacceptable descriptions only.  A much better description in terms of \redchi ($\chi^2=44.14$ with 43 \textsl{d.o.f.}) was obtained, when using the \texttt{bknpower} model, a model consisting of two power-laws with different photon indices $\Gamma_1$ and $\Gamma_2$, with the change from one photon index to the other taking place at a specific break energy $E_\text{break}$. It was not necessary to apply systematics for this fit. The model was attenuated by photo absorption at lower energies, expressed as equivalent column density of neutral hydrogen atoms $N_\mathrm{H}$. Data and best fit model are shown in Fig.~\ref{fig:spectrum}, and the parameters of the best fit model can be found in Tab.~\ref{tab:specpar}. The required new description of the model compared to the older spectra leads to an overall harder spectrum, as no exponential cut-off is visible anymore. The soft energy continuum has a slope which is consistent with previous published results, see, e.g., \citet{mattana06a}.

\begin{table}
\centering
\caption{Best fit parameters of the \texttt{bknpower} model.}
 \begin{tabular}{c|c|c|c|c}
$N_\mathrm{H}$ & $A_\text{bknpower}$ & $\Gamma_1$ & $\Gamma_2$ & $E_\text{break}$ \\
 $(6^{+4}_{-4})\times10^{22}$ & $0.22^{+0.16}_{-0.09}$ & $1.76^{+0.17}_{-0.17}$& $ 3.78^{+0.21}_{-0.18}$ & $27.9^{+1.2}_{-1.1}$\,keV
\end{tabular}
\label{tab:specpar}
\end{table}

\section{Outlook \& Conclusions}

We presented a timing and spectral analysis of \inte data of the large outburst of \ta in 2008 November. We confirm the unusually long spin period and measure a value of $\sim 5.3$\,h. During the outburst we found that the pulse period is drastically increasing, with a spin up rate as high as $\dot{P} = -2 \times 10^{-4}$\,\nicefrac{h}{h}. This rate is a magnitude higher than the one found by \citet{corbet08a} for the 2005 outburst. As these authors already stated, a change of this magnitude can not be due to the orbital motion, as it would lead to a mass-function for the companion of $>5\times10^6\msun$. On the other hand, if we assume that the neutron star is spinning close to equilibrium, i.e., that the Alfv\'en radius is equal to the Keplerian co-rotating radius, then a magnetic field of $\sim 10^{15}$\,G would be required. In the standard accretion picture of \citet{ghosh79a}, this assumption is justified as only close to equilibrium the torque can easily change direction. However, more sophisticated accretion models have been proposed for other sources, including retrograde spinning accretion disks \citep{perna06a} and spherical wind accretion. If no stable accretion disk forms and material and momentum is accreted from the stellar wind, a torque transfer of this magnitude might be possible (see Postnov et al. elsewhere in these proceedings). Continued monitoring of \ta is necessary in any case to clarify possible models and check if the source continues to spin down during quiet phases, while strongly spinning up during flares.

The pulse peak spectrum of \ta we presented is distinctly harder than the previously published spectra, but is otherwise featureless. Especially, we find no indication for a Cyclotron Resonant Scattering Feature (CRSF), as seen in many other X-ray binaries, where the neutron star possesses a magnetic field $\geq 10^{12}$\,G. Such a feature is the only direct way to measure the magnetic field strength in the X-ray producing region and could clarify the size of the accretion region. If \ta really has a magnetic field on the order of $10^{15}$\,G, it is not suprising that we do not see any CRSF, as the fundamental line energy would be $\geq 10$\,MeV, where the source is not visible anymore.

For a more detailed presentation of the results see a forthcoming publication by Marcu et al., 2011, in prep.


\bibliographystyle{PoS}
\begin{multicols}{2}

%

\end{multicols}

%

\end{document}